# MOMENT: a muon-decay medium-baseline neutrino beam facility

Jun Cao[1], Miao He[1], Zhi-Long Hou[1], Han-Tao Jing[1], Yu-Feng Li[1], Zhi-Hui Li[2], Ying-Peng Song[3,1], Jing-Yu Tang[1*], Yi-Fang Wang[1*], Qian-Fan Wu[1], Ye Yuan[1], Yang-Heng Zheng[4]

[1]Institute of High Energy Physics, CAS, Beijing 100049, China
[2]Sichuan University, Chengdu 610065, China
[3]University of Science and Technology of China, Hefei, Anhui 230029, China
[4]University of Chinese Academy of Sciences, Beijing 100049, China

**Abstract**: Neutrino beam with about 300 MeV in energy, high-flux and medium baseline is considered a rational choice for measuring CP violation before the more powerful Neutrino Factory to be built. Following this concept, a unique neutrino beam facility based on muon-decayed neutrinos is proposed. The facility adopts a continuous-wave proton linac of 1.5 GeV and 10 mA as the proton driver, which can deliver an extremely high beam power of 15 MW. Instead of pion-decayed neutrinos, unprecedentedly intense muon-decayed neutrinos are used for better background discrimination. The schematic design for the facility is presented here, including the proton driver, the assembly of a mercury-jet target and capture superconducting solenoids, a pion/muon beam transport line, a long muon decay channel of about 600 m and the detector concept. The physics prospects and the technical challenges are also discussed.

**Key words**: neutrino beam, CP violation, CW superconducting linac, mercury-jet target, muon decay channel, neutrino detection
PACS: 29.38.Db, 14.60.Pq, 29.20.Ej, 29.27.Eg

## 1. Introduction

### 1.1 Physics goals

In the last two decades, neutrino physics has made tremendous advancements, with the discoveries of neutrino oscillations and measurements on the mass and mixing parameters [1]. In 2012, the last mixing angle $\theta_{13}$ was discovered to be non-zero at Daya Bay and other neutrino facilities [1-7]. The unexpected large $\theta_{13}$, about 9°, changes the scenario of the neutrino oscillation study. It becomes easier to measure the mass hierarchy (i.e., the sign of $\Delta m^2_{31}$) and lepton CP violation phase. It is believed that the experiments to be carried out at the facilities either under construction or under planning, such as JUNO [8], PINGU [9], HyperK [10], and LBNE/LBNO [11-12] in the coming decade can determine the neutrino mass hierarchy and measure the oscillation parameters very precisely. Thus, new experimental facilities now shall focus on the measurement of the CP phase. The Neutrino Factory (NF) [13] which was first proposed fifteen years ago and has been undergoing continuous design optimization, is considered to be the ultimate facility to measure all the neutrino oscillation parameters. However, it has many technical problems [14] to be solved before it is materialized. Other possible accelerator-based neutrino beam facilities in Japan and Europe, namely, T2HK [15] and ESSnuSB [16-17], using neutrinos from pion decays instead of muon decays at neutrino factory, suffer from background issues and possibly, insufficient beam power.

---
[*] Corresponding authors: tangjy@ihep.ac.cn, yfwang@ihep.ac.cn

We propose here a dedicated facility, MOMENT (MuOn-decay MEdium baseline NeuTrino beam facility), for the CP phase measurement using neutrinos from muon decays. Some of the technically difficult issues at the neutrino factory, such as the muon cooling and muon acceleration are avoided. Neutrinos are thus produced only at low energies, namely in the range of 200-300 MeV. This is feasible since the sensitivity to CP phase is independent to the neutrino energy and it is free from $\pi^0$ background, as shown in Figure 1.

In this article, the design scheme for the MOMENT is presented.

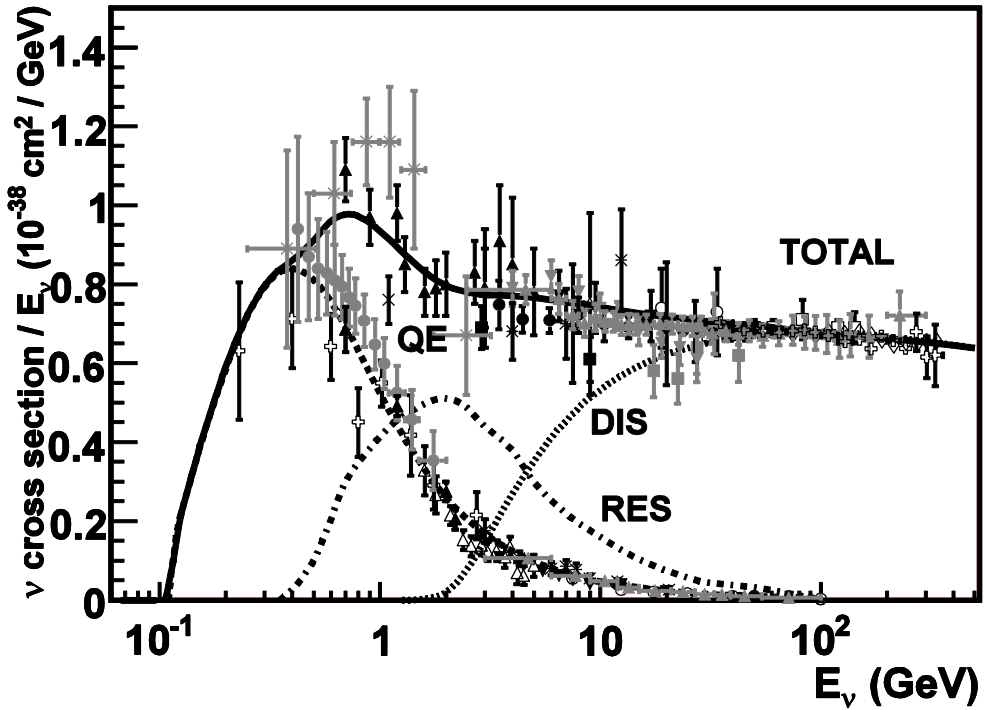

Figure 1: Neutrino charge current cross-sections as a function of neutrino energy [18-19]

## 1.2 Facility concept

To obtain a very intense neutrino flux for medium baseline neutrino oscillation experiments, a proton beam of very high beam power with an energy range of 1-3 GeV is mandatory, at least at the level of a few megawatts. Both pulsed and continuous-wave (CW) proton beams can be considered. A common consensus in the international accelerator community is that the highest beam power can only be provided by a CW proton linac which can deliver a proton beam in tens of MW. However, there still exist no such proton linacs in the world due to extremely challenging technological difficulties in designing and building them. Fortunately,

China is among the first countries which started the development of such high-power proton linacs for nuclear waste transmutation. The China Accelerator-Driven System (China-ADS) project is developing a 15-MW superconducting linac in the CW mode in multiple stages [20]. Thus it is reasonable that we adopt a similar proton linac as the proton driver for MOMENT.

For pion production which is the first step for a neutrino beam, the target which should withstand the 15-MW beam power is a critical issue. The pion production target should be a radially thin one to facilitate the emission of pions from inside the target. By now, no solid target of such definition has proven that it can survive with the bombardment of a 15-MW proton beam. Following the NF target design concept [14], we also adopt a mercury jet target. To collect pions efficiently, we should place the target in a very strong focusing field. With a CW proton beam, one cannot apply a focusing system based on magnetic horn as commonly used in many neutrino superbeams, since it can only work in a short time duration and with a low repetition rate. Here the concept of placing the target in a high-field superconducting (abbr. as SC) solenoid is adopted. The mercury jet target, the capture SC solenoid and the auxiliary devices constitute a very complicated target station. Compared with the pulsed proton beam at the NF, the CW proton beam at MOMENT helps a lot in easing heat mitigation and shock-wave problems in the target station, though the total heat deposit and the radiation dose rate due to the extremely high beam power are crucial problems to be solved here. As solenoid focusing does not distinguish between positive and negative pions which are produced simultaneously, the pion collection will include both types of pions. However, this will hinder the use of pion-decayed neutrinos for experiments, as the neutrinos generated by the opposite-charge pions will form unwanted background at the detector. The solution adopted here is to transport both types of pions and convert them into muons in a pion-decay channel, and then to separate the two muon beams in a special section.

Different from pion decays which happen mostly in a few tens meters, muons have much longer decay time. Therefore, the NF and the nuSTORM [21] have designed decay rings for muons to circulate many turns before they decay fully. However, with a CW proton beam as the primary one, the tertiary muon beam is almost a DC beam with a large momentum spread and transverse emittance. This hinders the storage of muons in a decay ring. Thanks to the relative low muon beam energy, we can design a long muon decay channel of about 600 m in which about 30% muons can decay, almost comparable to the decay efficiency in one long side of a racetrack decay ring. The selection of positive and negative muon beams is required before transporting the muon beam into the final decay channel. In addition, a bending section before the decay channel can help to get rid of wrong charged particles, pion-decayed neutrinos at the detector, to select the momentum range of the muon beam and to choose the direction of the neutrino beam.

The far detector is planned to be placed at a distance of approximately 100-150 km, the baseline with maximum oscillation probability for the average neutrino energy. A near detector can measure the neutrino spectra and beam profile. The detection of muon decayed neutrinos is quite different from pion-decayed neutrinos. It is immune from the intrinsic background as long as the detector can distinguish the lepton charges [13]. The pre-selected momentum range for the muon beam can limit the maximum neutrino energy to about 400-500 MeV, and this is also helpful in reducing the $\pi^0$ background. For the neutrino beam from the decay of a muon beam, we have electron neutrinos and muon anti-neutrinos in the decay channel but all the four species ($\nu_e$, anti-$\nu_e$, $\nu_\mu$, anti-$\nu_\mu$) at the far detector due to the oscillation. Therefore, we should design a detector capable of distinguishing the lepton charges, neutrino flavors and reject neutral current (NC) backgrounds from charge current (CC) interactions. Water Cherenkov (WC) detectors [22] have been proven to be effective in the flavor identification and CC/NC discrimination. Moreover, the neutron tagging technique in gadolinium (Gd) doped WC detectors [23], which is still being developed [24] by the Super-Kamiokande Collaboration, is superior to discriminate electron antineutrinos from electron neutrinos. Thus a large-volume Gd-doped WC detector may satisfy our requirement [25]. Of course, a magnetized sampling calorimeter can also satisfy the needs [14].

In summary, the proposed MOMENT facility mainly consists of a 15-MW superconducting linac in the CW mode as the proton driver, an assembly of the mercury jet target and the superconducting solenoids for pion production and collection, a pion/muon beam transport line, a long muon decay channel of about 600 m and a possible large-size detector of Gd-doped water. The schematic layout is shown in Figure 2. The neutrino source including the proton driver, target station and pion/muon transport channels can be hosted in the CSNS site [26], and a possible detector may be located at the JUNO site. The distance between the two sites is about 150 km, consistent to the average neutrino energy of 200-300 MeV.

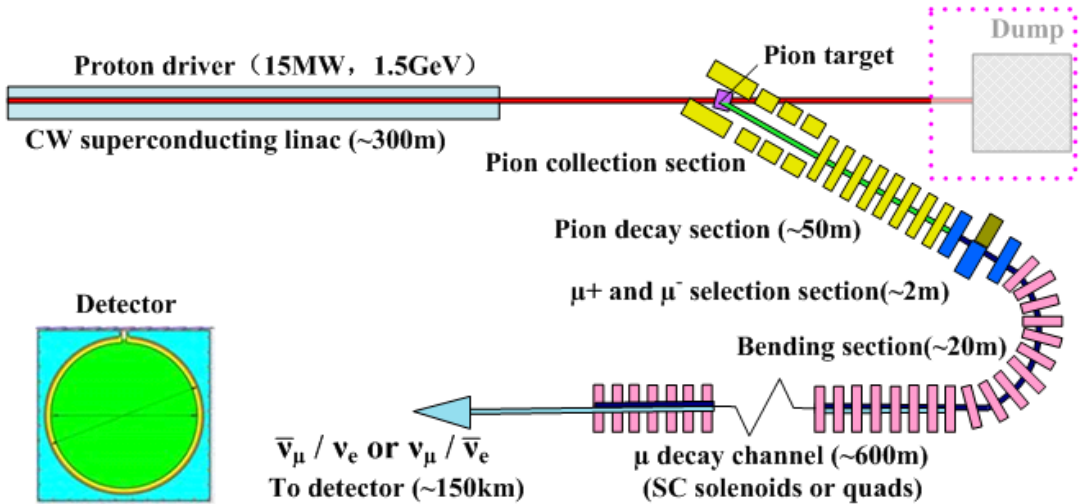

Figure 2: Schematic layout of the MOMENT facility

## 2. Proton driver

### 2.1 Introduction

The proton driver is a CW superconducting linac with a beam power of 15 MW, and the beam energy is still in optimization with a range of 1.5-2.5 GeV depending on the efficiency of neutrino production and the cost. The nominal design energy is 1.5 GeV, the same as for the China-ADS linac [20], and this corresponds to a proton-on-target (POT) per year of $1.1\times10^{24}$ POT/y. As there are a lot of technical challenges in building such a high beam power proton linac, we will profit from the China-ADS project and other CW linac projects which are developing the relevant techniques. For example, in about 2020, the China-ADS project is expected to complete an experimental facility with a linac of 250 MeV working in the CW mode. All the SC cavity types needed for MOMENT except the high-beta one will have been tested with beam by then. As MOMENT is a scientific research facility, it is not necessary for the driver linac to have the extremely high reliability as required by ADS, and this will reduce the technical difficulties greatly and the cost significantly. We can simplify the design by taking away many installed redundancies.

**2.2 Lattice structure and beam dynamics simulation results**

The main difference between the MOMENT linac and the China-ADS linac are: the former has only one front-end instead of two parallel ones for the latter; the former uses full applicable RF voltage as the nominal setting instead of about one third reservation for the local compensation in the cases of cavity failures for the latter. This means that the MOMENT driver linac tolerates more frequent beam trips of longer duration than the China-ADS driver linac. The reliability of the MOMENT linac can be still guaranteed by applying the global compensation method as at SNS when some SC cavities fail [27]. The schematic layout of the MOMENT linac is shown in Figure 3. In the following, some more details about different sections of the linac and beam dynamic studies are given.

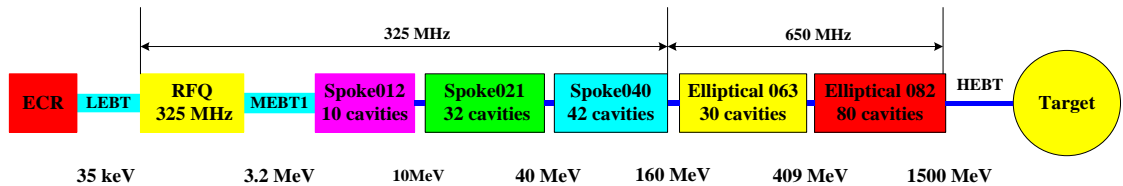

Figure 3: Schematic layout of the MOMENT driver linac

The lattice structures for different sections are shown in Figures 4 and 5. The front end is defined as the low-energy section including the ion source, LEBT, RFQ, MEBT, and Spoke012 section with the output energy of 10 MeV, which is totally the same as Injector Scheme I of the China-ADS linac. This is the most crucial part for a CW linac, where both the room-temperature RFQ, the very low-beta spoke cavities and the long cryomodule are difficult to develop. The test stand for the front-end is under construction at IHEP.

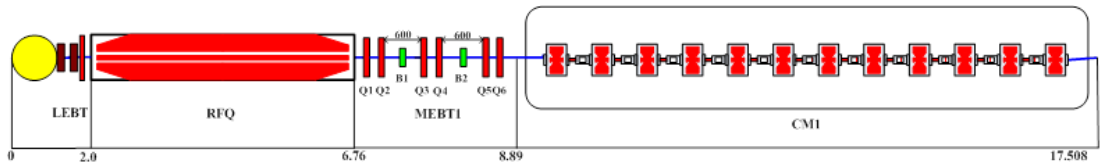

Figure 4: Layout of the front-end of the driver linac

For the main linac part, there are five sections in four SC cavity types, and the sections are mainly determined by the focusing structures. As this is a high-power linac, the criteria of 1 W/m beam loss rate along the accelerator [28-29] should be applied. The lattice structures should meet the requirements of the zero-current phase advance per period less than 90° degrees and smooth changes in the phase advance per meter between neighboring sections. The main parameters of the main linac are summarized in Table 1.

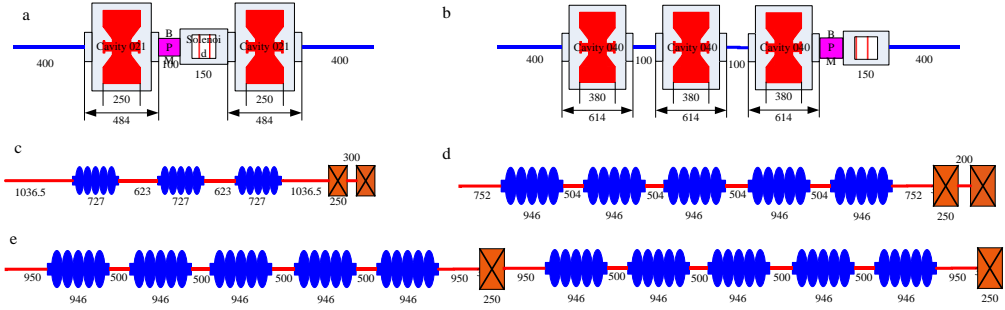

Figure 5: Lattice structures of the main linac (a: Spoke021 section - RSR, b: Spoke040 section - $SR^3$, c: Ellip063 section - $FDR^3$, d: Ellip082 - $FDR^5$, e: Ellip082 section - $FR^5DR^5$; S for SC solenoids, R for SC resonators, F/D for warm quadrupoles)

Table 1: Main parameters of the proton driver linac

|  | Spoke012 | Spoke021 | Spoke040 | Ellip063 | Ellip082-1 | Ellip082-2 | Total |
|---|---|---|---|---|---|---|---|
| Energy (MeV) | 10 | 40 | 160 | 409 | 1000 | 1500 | 1500 |
| Cavity number | 12 | 32 | 42 | 30 | 45 | 35 | 196 |
| Focusing structure | RS | RSR | SR3 | FDR3 | FDR5 | FR5DR5 |  |
| Total length (m) | 8.768 | 43.456 | 87.444 | 150.444 | 230.994 | 293.544 | 293.544 |
| Section length (m) | 8.768 | 34.688 | 43.988 | 63.000 | 80.550 | 62.550 |  |
| Cryomodules | 1 | 8 | 7 | 10 | 9 | 7 | 42 |
| Synchronous phase | -43–-30 | -45–-30 | -25 | -20 | -15 | -15 |  |

The synchronous phase increases along the linac to obtain higher acceleration rate as the bunch length shrinks with the increasing energy, but it is important to have a large ratio of longitudinal acceptance to bunch area.

Multi-particle simulations have been performed to verify the validation of the design. The RFQ is simulated by ParmteqM [30] and a truncated-4σ Gaussian distribution with 100,000 particles with the TraceWin code [31] is used for the simulations of the rest linac. For all the SC cavities, the 3-D field maps based on the cavity electromagnetic designs are used. For the transverse elements, the hard-edge approximation is used. Figures 6 and 7 show the rms envelopes and rms emittances along the linac. We can see that the envelopes change smoothly and the emittance growths are small.

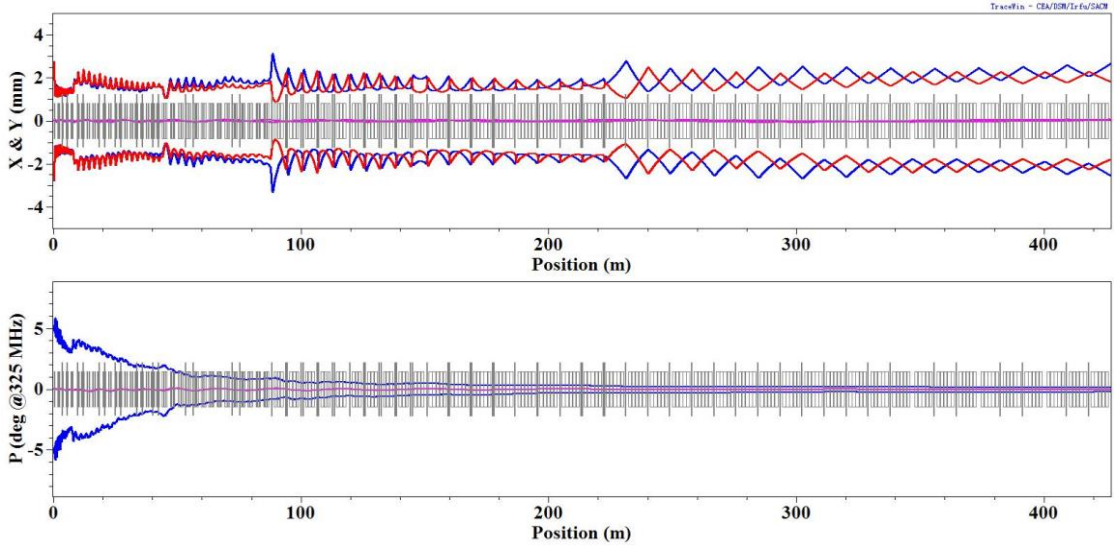
Figure 6: The rms beam envelopes along the linac

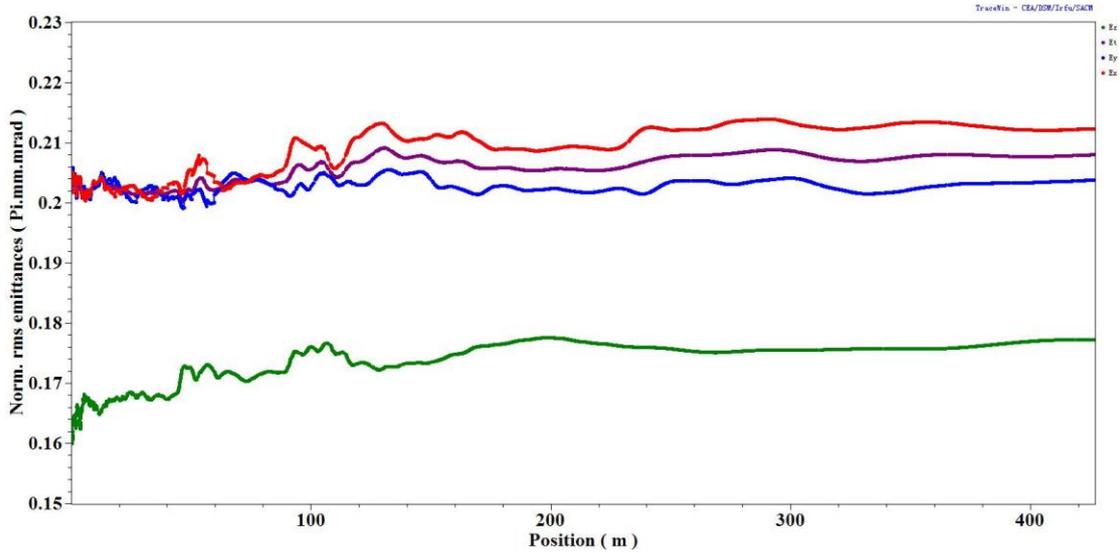
Figure 7: Evolutions of the rms emittances along the linac

**2.3 Alternative design**

As mentioned above, the proton energy may be optimized according to the global optimization between the neutrino production and the facility construction cost. Even for higher proton energy, the total beam power will be kept at 15 MW. This means that with higher energy the beam current will be lower which benefits the beam dynamics to some extent. The alternative design for extending the linac has been studied. A straightforward solution is just to add more Ellip082 cavities to cover the energy range from 1.5 GeV to 2.0 or 2.5 GeV. Another solution is to use a new high-beta elliptical cavity type (Ellip093) to cover the high-energy part of 1.0 to 2.5 GeV. The study shows that with the latter one can save 15 cavities and the linac length of about 20 meters, which is considered only a modest gain that may not compensate for the additional effort on developing the new cavity type.

Following the progress of the China-ADS project, there will be more and more inputs from the hardware developments to be integrated in the design and the design will be further optimized.

**2.4 High-energy beam transport line**
The proton beam from the linac will be guided to impinge the target horizontally. Although this part can be flexibly designed to meet the general layout, an achromatic bending section is required to avoid the beam wobbling at the target and the back-streaming neutrons towards the linac tunnel. The final section will focus the proton onto a small spot of about 4-6 mm in diameter at the mercury jet target.

Although it is very difficult, we will continue the study to extract the used primary proton beam to an external beam dump instead of dumping it in the mercury pool as the IDS-NF does [14]. If this is possible, the dump beam line and the dump station are still technically difficult due to poor proton beam quality and very high beam power which is about a few megawatts.

## 3. Target and secondary/tertiary particle collection

**3.1 Mercury jet target**
As mentioned in Section 1.2, the pion production target should be a radially thin one to facilitate the emission from the target of the pions generated by the proton-target interaction and has to withstand the bombardment of a 15-MW power proton beam. For now it is believed that only a mercury jet target can meet the requirement, so it is adopted for the baseline design of the MOMENT target, just similar to the NF target design. The MERIT collaboration has demonstrated by experiments that such a mercury jet target is technically feasible [32]. The experiments also show that a very high-field helps the confinement of the jet during the beam impinging period. Compared with the NF target or MERIT target, the peak beam power at MOMENT is much lower, so there will be no such problems as jet confinement and shockwave. Other options are also to be considered. For example, fluidized tungsten powder which is under study by RaDATE [33], can be considered if the feasibility is demonstrated in the future.

**3.2 Pion production and collection**
Different proton beam energies from 1.5 to 2.5 GeV have been used in the simulations to find a solution which is a trade-off among the technical difficulties, cost and neutrino production. At the moment, although the higher the proton beam energy the higher pion production yield, it is considered that the gain by increasing the proton beam energy is marginal if one keeps the beam power unchanged. The baseline design is with a 1.5-GeV proton beam, but it can be modified later.

The experience of high-field SC solenoid for pion capture from MUSIC [34] and COMET [35-36] is used here in the design. The capture, decay and transport of secondary pions and tertiary muons are simulated mainly by using the G4Beamline code [37]. It is found that the magnetic field level has a strong impact on the pion collection efficiency, especially on the core emittance of the pion beam. Different capture fields have been studied, and the field level of about 10 T or higher is considered necessary to efficiently capture the pions. However, the field level in this very large aperture magnet is limited by the superconducting magnet technology [38-39]. The baseline design adopts 14 T for the central field in the target region, similar to the design magnetic field by superconducting coils in the Neutrino Factory [14]. The use of high-TC magnet may permit operation at a higher field and will be studied in the future. Due to the production of very high neutron and gamma fluxes by the proton-target interaction, the heat deposit and irradiation effect in the SC coils are crucial problems for the cryogenic system and the quenching of superconductivity. A well-designed shielding to absorb the neutrons and gamma rays is mandatory, but it will increase the dimensions of the SC solenoid largely which has a very important impact on the magnetic field distribution and the cost. Figure 8 shows the schematic of the target station, which is somewhat similar to the NF target station [14].

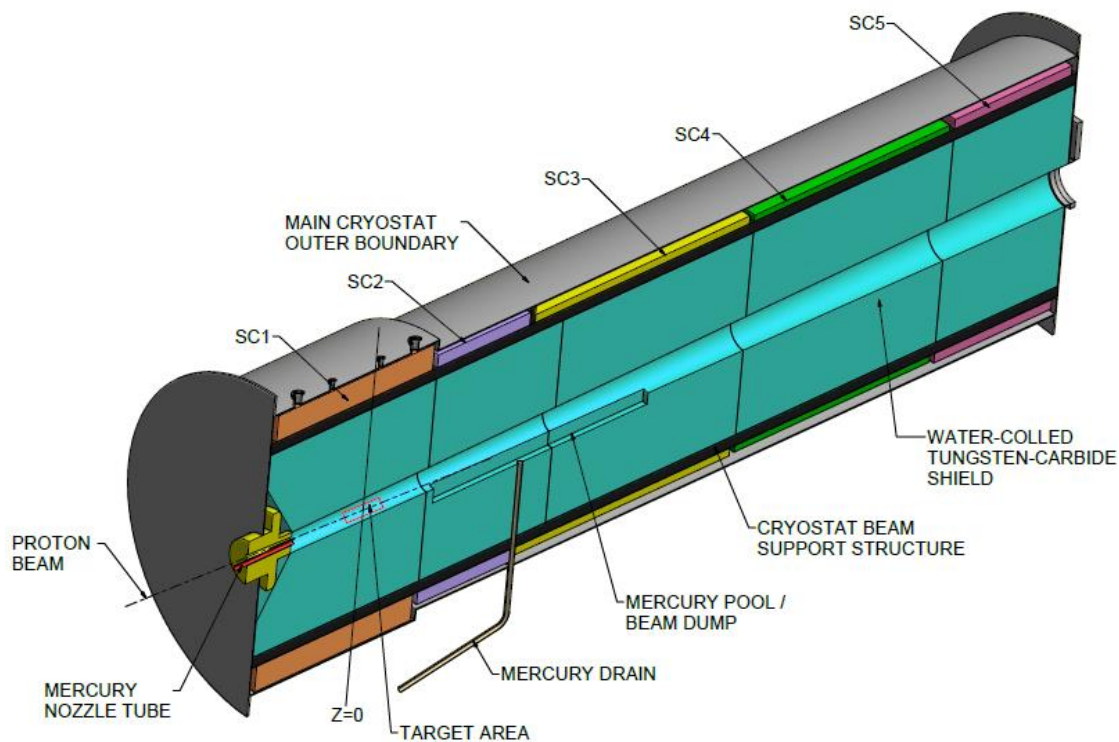

Figure 8: Schematic of the MOMENT target station

The pions emitted from the target, together with the muons by either stopping pions or shortly decayed in-flight pions, are captured with a high efficiency. As low energy muons do not contribute to the neutrino oscillation experiment, and they will be discarded in the downstream beam transport line. The thickness and length of the target are for the optimization of the pion production and capture. For the moment, the mercury jet target is designed to have a radius of 5 mm and an effective length of 300 mm. The proton beam spot is 1 mm in rms. The mercury jet has a small slanted angle with respect to the proton beam, which is to be optimized. The calculated pion production rate is 0.10 pion/proton. With the capture field of 14 T, the collection efficiency of the forward pions/muons is about 44%. A field tapering section by five SC solenoids is used to adiabatically decrease the field level from 14 T in the target region to 3-4 T in the decay/transport channels, as shown in Figure 9. This process is important to convert the transverse momentum into the longitudinal momentum for better beam transport in the downstream beam transport lines. Figure 10 shows the momentum spectrum of the extracted pions. As a comparison, two other capture field levels of 7 T and 10 T have also been studied.

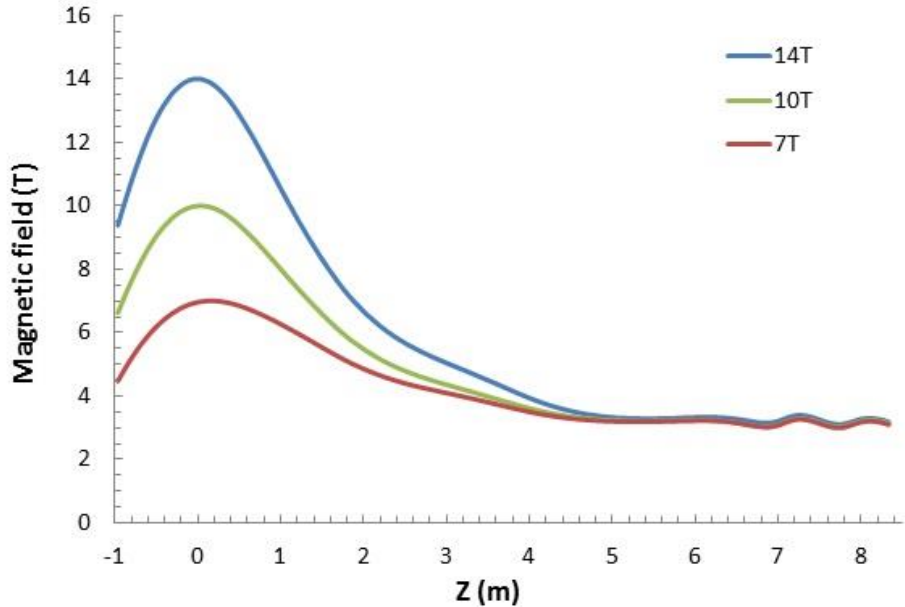

Figure 9: Magnetic field pattern on axis along the field tapering section

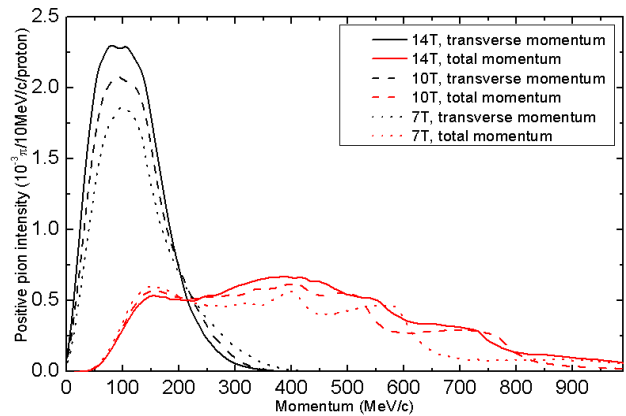

Figure 10: Momentum spectrum of the extracted pions after field tapering with 7 T, 10 T and 14 T capture fields, respectively

**3.3 Heat deposit and radiation dose rate in the capture solenoid**

Nb$_3$Sn superconducting coils [38] will be used for the capture solenoid. The main technological challenges are the high heat deposit and radiation dose rate in the SC coils by neutrons and gamma rays emitted from the target. A shield structure by tungsten between the target and the SC coils is crucial in reducing the two parameters. FLUKA [40-41] simulations have been carried out to optimize the shielding design. The simulations show that the shielding can absorb almost all the gammas and the majority of neutrons. The heat deposit density is shown in Figure 11. If we want to limit the energy deposited in the SC coils below 1 kW which is considered already a very heavy burden for the cryogenic system, the shielding thickness at the target location should be at least 800 mm. This defines the inner radius of the SC coils as 1050 mm. The inner bore of the shielding block is tapered to keep it away from the used protons and also the high-energy pions. The minimum inner radius of the shielding block at the target area is 200 mm.

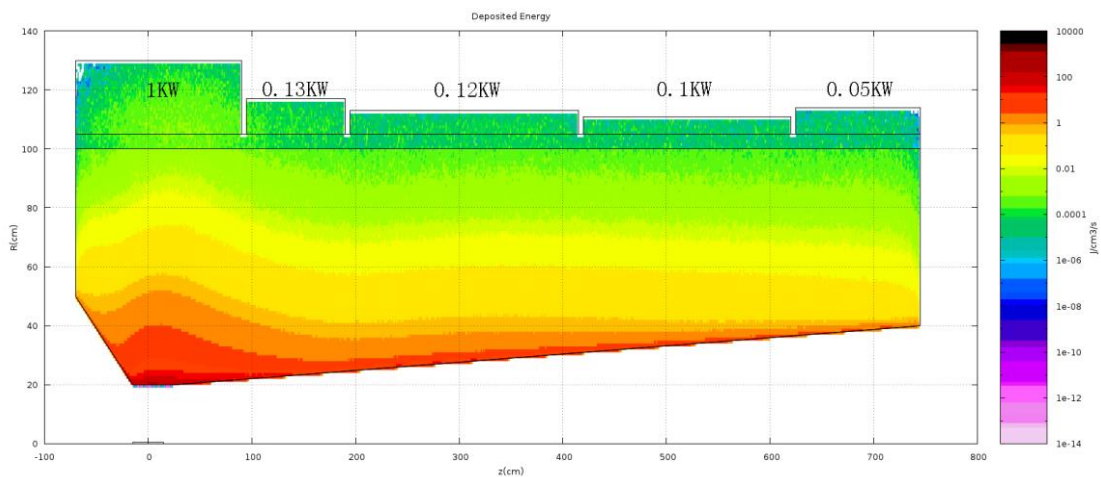

Figure 11: Heat deposition in the shielding block and the SC coils

**3.4 Treatment of used protons**

Due to extremely high beam power, the proton beam poses a very challenging problem for the design of the target station. Only a small part of the primary beam power is converted into pion production and consumed by other nuclear reactions, and an important part is lost in the target by the ionization process. There will be two different proton compositions after the interaction between the primary beam and the target, one is the degraded primary proton beam which has lost a part of its energy or power mainly due to ionization, and the other is the scattered protons from the target during the interaction which spread over a large momentum range.

It will be very interesting to guide the degraded proton beam to an outer beam dump instead of sending it into the local mercury pool directly, so the neutron production yield in the target station can be largely reduced. This is important for designing the SC coils and the shielding block. However, the degraded beam quality and the sophisticated structure of the target station may hinder its feasibility in the engineering design.

The scattered protons will come out from the target area together with the pions. A part of them will be intercepted by an absorber, and others will go out into the downstream pion decay channel and heat up the SC solenoids and the beam pipes in the channel, which is to be discussed in Section 4.

In the future, lot of efforts will be needed to study how to mitigate the beam power of the used protons.

## 4. Muon beam transport and decay channels

**4.1 Pion decay channel**

Thanks to the almost identical moment for both a pion and its descendent muon, both pion and muon beams can be transported in the same focusing channel when pions continue to decay into muons. In order to avoid the background by the neutrinos from pion decays at the detector, the muon beam line is designed to have a transport section where pions continue to decay, a bending section where very high-momentum pions are eliminated and a decay section where the probably selected muons decay to produce the required neutrino beam. Because our goal is to use neutrinos with the averaged energy close to 300 MeV, the required muon beam should have momentum close 300 MeV/c. Therefore, the average momentum of 300 MeV/c and the spread of about ±50% are chosen for the muons in the decay section. This means that other pions or muons outside the moment range which occupy more than 50% of the total collected pions from the target/collection region, as shown in Figure 12, should be eliminated by the bending section.

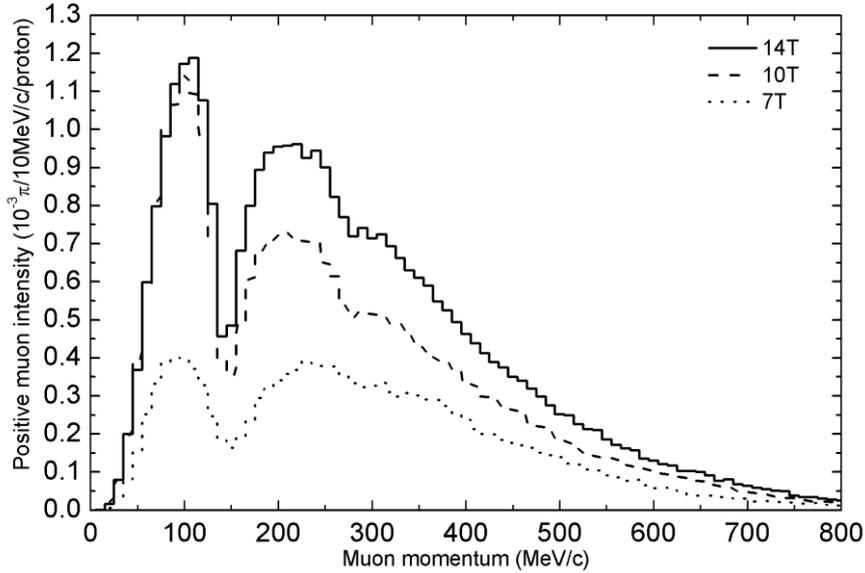

Figure 12: Muon momentum spectrum at the end of the pion decay channel

Studies show that with a very large transverse emittance, large momentum range and low average momentum, among different focusing structures SC solenoids are considered the best for the transverse focusing. The transverse acceptance and momentum acceptance along the pion/muon transport lines are very large in SC solenoids-based focusing channels, and determined mainly by the requirement on the muon beam quality in the muon decay channel, which are in the orders of 100 $\pi$mm.rad and ±50%, respectively. Some details about the transport beam lines are shown in Table 2. The transverse acceptance for a different momentum other than the reference momentum is inversely proportional to the momentum.

Table 2: Transverse acceptances of the pion/muon transport lines

|  | Ref. momen. (GeV/c) | Aperture (mm) | Peak field (T) | Acceptance ($\pi$mm.rad) | Trans. eff. |
|---|---|---|---|---|---|
| Pion decay channel | 320 | 600 | 3.7 | 141 | 68% |
| Bending section | 300 | 600 | 3.7 | 100 | 60% |
| Muon decay channel | 300 | 800 | 1.0 | 65 | 76% |

A pion decay channel with a length of about 50 m is designed to convert more-than 90% pions into muons. Because there are also many scattered protons from the proton-target interaction coming out together with the pion/muon beam, which possess high beam power, a well-designed shielding is required to mitigate the heat to avoid the heat-up in the SC solenoids. A similar design such as the chicane-type adopted by the Neutrino Factory can be applied here. In addition, the beam power related to the pion/muon beam also reaches up to about 100 kW, which should be also treated carefully along the channel.

Except in the beginning part where the focusing strength is matched to the capture field in the target region and the curved solenoids are for eliminating protons, the long straight section of the channel is made of a pure FOFO focusing structure by SC solenoids. The averaged magnetic field for the SC solenoids is about 3.4 T and the aperture for the beam pipe is about 600 mm.

### 4.2 Selection of positive and negative muons

Both positive and negative pions produced at the target are transported without preference by a solenoid-based focusing channel, and it is the same for the positive and negative muons. However, only one of the positive and negative muon beams is used for neutrino production at a time, so one of them should be removed from the main channel before entering the muon decay channel. Because the transverse emittance is very large here, it is difficult to use a simple chicane-type structure with usual dipole magnets. We have adopted a design using a group of superconducting dipole magnets which have also high field gradients to form a strong triplet-like focusing. Figure 13 shows the schematic of the $\mu^+/\mu^-$ selection. The discarded muon beam, either positive or negative, is so strong that it far outweighs the existing and proposed muon beams. It would be interesting to guide it to a dedicated muon facility for muon-based research, such as muon physics and μSR (Muon Spin Rotation, Relaxation and Resonance) applications.

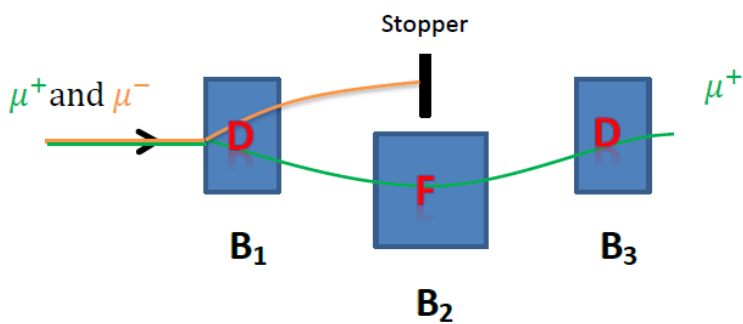

Figure 13: Schematic of the $\mu^+/\mu^-$ selection

### 4.3 Bending section

As mentioned above, the bending section before the muon decay channel can help get rid of the pion-decayed neutrinos at the detector when needed and select the momentum range of the muon beam. This part can be flexibly adapted to the facility layout design with its bending angle and length. This conceptual design uses a bending angle of 90° with a length of 27 m. The bending is made by slanted SC solenoids, which effectiveness has been approved at MuSIC [42]. The slanted angle for each solenoid has an important impact on the beam centroid excursion. We need many short solenoids with a small slanted angle such as 2° used here. In order to reduce the coupling between the transverse phase planes, it is necessary to use alternate fields for the neighboring solenoids. Figure 14 shows the field distribution and the beam centroid excursions along the bending section.

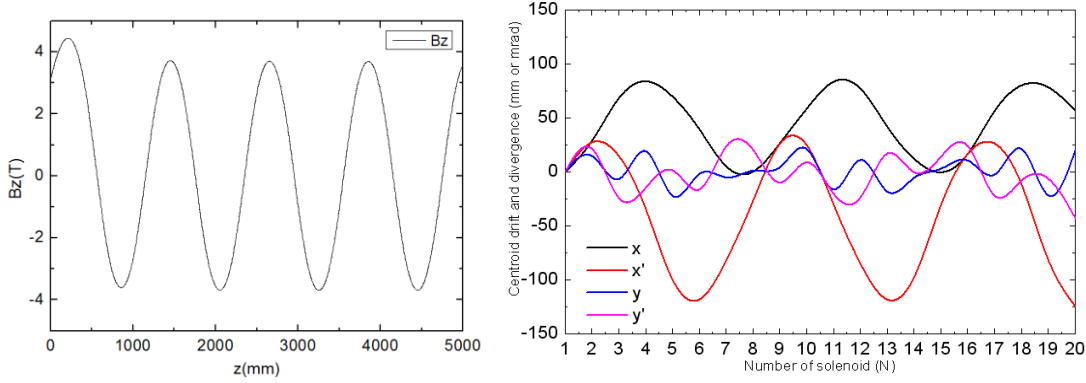

Figure 14: Field distribution (left) and the beam centroid excursions (right) in a part of the bending section

There are some quite energetic pions which are still alive after passing through the pion decay section. They can continue to decay in the bending section. However, for those with high kinetic energy they may survive until the entrance of the downstream muon decay channel, and they may be not desired by the detector. Therefore, it is important to limit the momentum acceptance of the bending section to remove the high-energy pions. This can be assisted by applying a small dipole field to the solenoids by attaching special coils to the main coils. The dipole field is also helpful to reduce the beam centroid excursion or the aperture of the beam pipe which is strongly related to the cost of the channel.

**4.4 Muon decay channel**
The long decay channel of 600 m in length is to obtain a decay probability of about 33% for the average muon momentum of 300 MeV/c. A matching section by decreasing the solenoid field from 3.7 T to 1.0 T is to adiabatically convert partial transverse momentum to longitudinal momentum, which is similar to the matching section in the pion decay channel. This is very important to reduce the transverse divergent angle for the muon beam which is critical to obtain a required neutrino spectrum with an average energy larger than 200 MeV at a detector of 150 km in distance, because high-energy neutrino is emitted only within a very small solid angle with respect to the muon's direction. Figure 15 shows the neutrino energy dependence on the off-axis angle and momentum of a muon. The FOFO focusing channel has a transverse acceptance of 65 πmm.rad for the reference momentum with a beam pipe of 800 mm in diameter. The total muon beam intensity in the decay channel is $1.0\times10^{15}$ $\mu^+$/s or $1.8\times10^{22}$ $\mu^+$/y, and the neutrino yield (in pair) is $5.4\times10^{21}$ ν/y which is more than twice the one at the Neutrino Factory. The neutrino spectra at the detector with simulated particles are shown in Figure 16. The averaged energy for muon anti-neutrinos is about 240 MeV, and the neutrino flux is $4.7\times10^{11}$ ν/m$^2$/y at the detector. The neutrino fluxes for different capture fields are summarized in Table 3.
Similar to the muon beam discarded by the muon beam selection section, the remaining undecayed muons at the end of the muon decay channel can also be used for other muon applications.

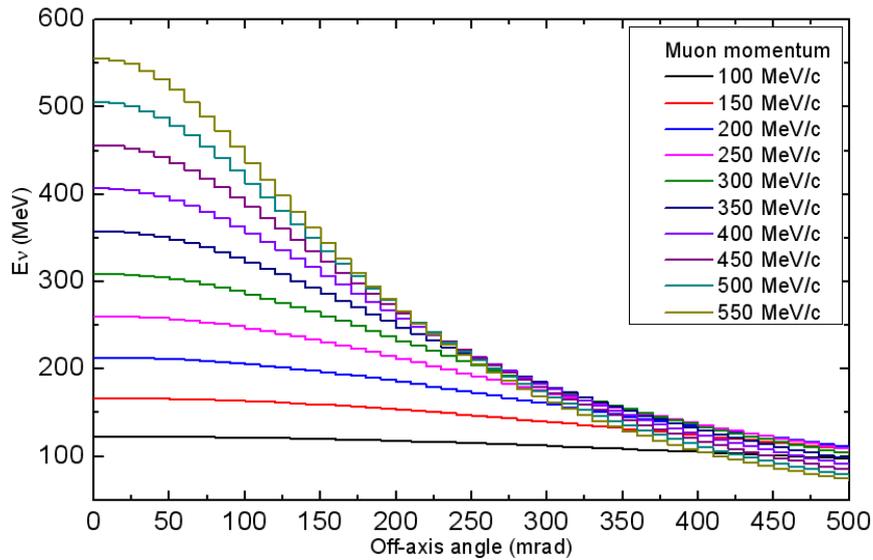

Figure 15:   Neutrino energy dependence on the off-axis angle and momentum of a muon

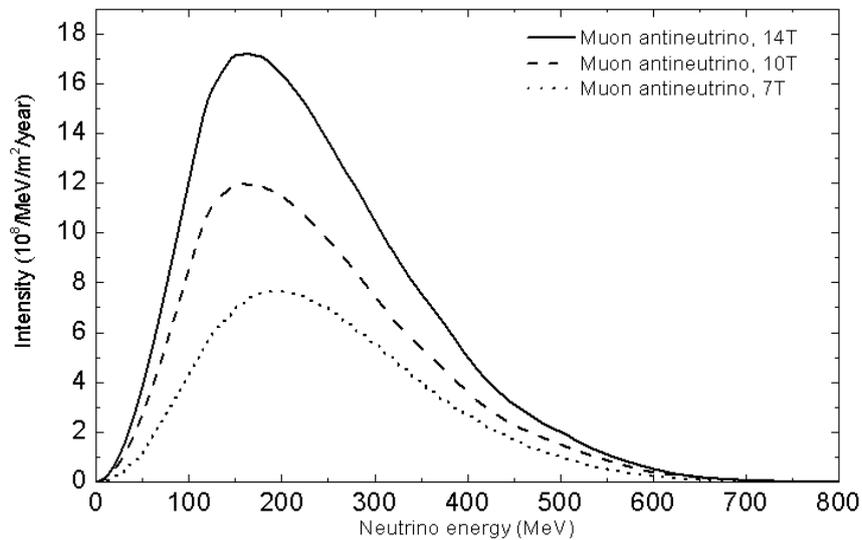

Figure 16: Muon anti-neutrino energy spectra at the detector

Table 3: Neutrino fluxes at the far detector for different capture fields at the target

| Field level | Neutrino flux (ν/m$^2$/y) |
|---|---|
| 7 T field | $2.1\times10^{11}$ |
| 10 T field | $3.3\times10^{11}$ |
| 14 T field | $4.7\times10^{11}$ |

## 5. Detector concept

A muon decay produces two neutrinos of different lepton charges and flavors, namely, a muon anti-neutrino and an electron neutrino for a $\mu^+$ decay, a muon neutrino and an electron anti-neutrino for a $\mu^-$ decay. Thus all the four neutrino flavors are present and the detector should be able to distinguish the charge and the flavor of neutrinos. It should also be able to distinguish charge current interactions (CC) from neutral current backgrounds, which could be very small in the case of low energy neutrino beams.

As mentioned in Section 1.2, a Gd-doped WC detector should be suitable for our purpose [25]. From the experience of the Super-Kamiokande experiment, water Cerenkov detector can well identify electrons from muons, and reject NC backgrounds from CC signal. A Gd-doped WC detector can well identify electron anti-neutrinos by inverse beta-decays while electron neutrinos by scattering-type of signal. The magnetized iron neutrino detector (MIND) or the magnetized liquid Argon (LAr) detector, which are being studied for the Neutrino Factory [14], is also well suitable. There is plenty of room for imagination in this direction, and novel detector concepts to match the special neutrino beam of MOMENT will be investigated. The target mass is expected to be sub-million tons to quickly accumulate statistics.

## 6. Discussion and conclusions

Although the scheme looks no stop sign in the moment, there are many technical challenges to be solved by R&D efforts in the coming years and during the engineering construction. The most challenging one is with the proton driver and the target station. For the proton driver, there are issues like RFQ working in the CW operation mode, low-beta superconducting cavities, high-power and large-scale RF amplifiers, cryomodules with many elements and high average heat load, very strict beam loss control etc. For the target station, there are problems related to mercury jet and cooling circulation, high-field SC solenoids, shielding design, very high radiation dose rate and very high heat load in the solenoids, treatment of the primary and scattered proton beams etc. The China-ADS project is executing a strong R&D effort and will build experimental facilities to solve the design and technical issues related to a high-power CW proton linac. The neutrino beam facility will profit from its outputs. The target station and the pion decay channel at MOMENT share many technical issues with the Neutrino Factory. A strong collaboration will be established between the two studies. Concerning the detector, which is still under the first investigation, the related technical issues are still to be identified.

The study on a muon-decayed medium-baseline neutrino beam facility presented above is still preliminary, more detailed studies and optimizations are to be carried out in the future. The extremely high neutrino flux and properly-defined neutrino energy spectra make it very competitive in measuring the CP violation phase when other major neutrino oscillation parameters are determined. In addition, the facility provides an extremely intense muon beam of $10^{14}$/s in parallel to the neutrino beam, which can be exploited for muon physics or other researches based on muon sources. The principle of producing intense and low-energy neutrino beam based muon decays is general and can be also applicable to other facilities.

## Acknowledgements

The authors wish to express their thanks to the international neutrino beam community who welcomes the proposal and gives many valuable suggestions. The discussions within the Chinese high-energy physics community are also encouraging the pre-conceptual design. Thanks also go to Huayan He for making mechanical drawings. The study was supported jointly by National Natural Science Foundation of China (Projects 11235012, 11335009, and 10875099) and the CAS Strategic Priority Research Programs-JUNO and China-ADS.

## References:

[1] Y.F. Wang, Proc. of IPAC2013, Shanghai, China, (2013) p.4010
[2] Daya Bay Collaboration, (F.P. An *et al.*), Phys. Rev. Lett. **108**, 171803 (2012).
[3] Daya Bay Collaboration, (F.P. An *et al.*), Chin. Phys. C **37**, 011001 (2013).
[4] Double Chooz Collaboration, (Y. Abe *et al.*), Phys. Rev. Lett. **108**, 131801 (2012).
[5] RENO Collaboration, (J.K. Ahn *et al.*), Phys. Rev. Lett. **108**, 191802 (2012).
[6] T2K Collaboration, (K. Abe *et al.*), Phys. Rev. Lett. 107, 041801 (2011).
[7] MINOS Collaboration, (P. Adamson *et al.*), Phys. Rev. Lett. 107, 181802 (2011).
[8] Y.F. Li, J. Cao, Y.F. Wang and L. Zhan, Phys.Rev. D **88**, 013008 (2013).
[9] D. J. Koskinen, Mod. Phys. Lett. A **26**, 2899 (2011).
[10] K. Abe *et al.*, (The HyperK Collaboration), arXiv:1109.3262.
[11] T. Akiri *et al.*, (The LBNE Collaboration), arXiv:1110.6249.
[12] S. Bertolucci *et al.*, arXiv:1208.0512.
[13] S. Geer, Phys. Rev. D **57**, 6989 (1998).
[14] R. J. Abrams *et al.* (IDS-NF Collaboration), Interim Design Report No. CERN-ATS-2011-216; arXiv:1112.2853.
[15] T. Ishida for Hyper-K working group, arViv:1311.5287
[16] E. Baussan et al., (ESSnu Collaboration), ArXiv: 1309.7022v3 [hep-ex], 2013
[17] T.R. Edgecock *et al.*, Phys. Rev. ST Accel. Beams **16**, 021002 (2013).
[18] J.A. Formaggio and G.P. Zeller, Rev. Mod. Phys. **84**, 1307 (2012).
[19] J.A. Formaggio, G.P. Zeller, arXiv: 1305.7513
[20] Zhihui Li et al., Phys. Rev. ST Accel. Beams 16, 080101 (2013)
[21] P. Kyberd *et al.*, arXiv:1206.0294.
[22] Super-Kamiokande Collaboration (S. Fukuda et al.), Nucl. Instr. and Meth. **A501**, 418 (2003).
[23] J. F. Beacom and M. R. Vagins, Phys. Rev. Lett. **93**, 171101 (2004).
[24] Super-Kamiokande Collaboration (H. Watanabe *et al.*), Astropart. Phys. **31**, 320 (2009) / T2K Collaboration (K. Abe *et al.*), Phys.Rev. D **88**, 032002 (2013).
[25] The detector concept and simulation results will be presented in a separate paper elsewhere.

[26] J. Wei, H.S. Chen, Y.W. Chen et al., Nucl. Instr. and Meth., A600 (2009) 10-13

[27] J. Galambos et al., Proc. of the Fifth International Workshop on the Utilisation and Reliability of High Power Proton Accelerators, Mol, Belgium, 2007.

[28] T.P. Wangler, Principles of RF Linear Accelerators, 1998, Wiley, New York

[29] N.V. Mokhov, and W. Chou (Eds.), Proc. of Workshop on Beam Halo and Scraping, Lake Como, (1999)

[30] http://laacg1.lanl.gov/laacg/services/download_sf.phtml

[31] http://irfu.cea.fr/Sacm/logiciels/index3.php.

[32] K.T. McDonald et al., Proc. of PAC2009, Vancouver, BC, Canada, (2009) p.795

[33] http://www-radiate.fnal.gov/index.html

[34] MuSIC Collaboration, The MuSIC project under the center of excellence of sub atomic physics, 2010. (http://133.1.141.121/~sato/music/doc/MuSIC.pdf)

[35] Yoshitaka Kuno, Nucl. Phys. Proc. Suppl., 225-227 (2012) 228-231

[36] COMET Collaboration (Y.G. Cui et al.), KEK-2009-10, June 2009

[37] T. J. Roberts et al., Proceedings of 2011 PAC, New York, USA, (2011) 373-375

[38] A.V. Zlobin, IEEE/CSC & ESAS EUROPEAN SUPERCONDUCTIVITY NEWS FORUM (ESNF), No. 16, April 2011

[39] P. Loveridge, in: Solenoid capture workshop, November 29-30, 2010

[40] G. Battistoni et al., Proc. of the Hadronic Shower Simulation Workshop 2006, Fermilab, M. Albrow, R. Raja eds., AIP Conference Proceeding 896, 31-49, (2007)

[41] A. Ferrari, P.R. Sala, A. Fasso, and J. Ranft, FLUKA: a multi-particle transport code, CERN-2005-10 (2005), INFN/TC_05/11, SLAC-R-773

[42] A.Sato, Y. Kuno, H. Sakamoto et al., Proc. of IPAC2011, San Sebastian, (2011) p.820